\documentclass[10pt]{article}

\usepackage{amsmath}
\usepackage{xcolor}
\usepackage{microtype}
\usepackage{graphicx}
\usepackage{multirow}
\usepackage{booktabs} 
\usepackage{hyperref, stmaryrd}
\usepackage{geometry}[margin = 1cm]

\usepackage[french]{babel}
\usepackage[utf8]{inputenc} 
\usepackage[T1]{fontenc}    
\usepackage{amsfonts}       
\usepackage[font=scriptsize]{caption}

\usepackage{nicefrac}       
\usepackage{microtype}      
\usepackage{amssymb}
\usepackage{amsthm}
\usepackage{dsfont}
\usepackage{comment}
\usepackage{bbold}

\newcommand{\syndrome}{\text{syndrome}}
\newcommand{\coutErranceParJour}{\textit{coutErranceParJour}}
\newcommand{\tempsErrance}{\color{blue} \text{tempsErrance}}
\newcommand{\coutMedecinSpeMR}{\textit{coutMedecinSpeMR}}
\newcommand{\tempsMoyenErrance}{\textit{tempsMoyenErrance}}
\newcommand{\coutMedecinNonSpeMR}{\textit{coutMedecinNonSpeMR}}
\newcommand{\nbMoyenMedecinConsultes}{\textit{nbMoyenMedecinConsultes}}

\DeclareMathOperator{\cout}{\textbf{coût}}

\usepackage{setspace}
\makeatletter
   \def\vhrulefill#1{\leavevmode\leaders\hrule\@height#1\hfill \kern\z@}
\makeatother

\newcommand*{\vcenteredhbox}[1]{\begingroup
\setbox0=\hbox{#1}\parbox{\wd0}{\box0}\endgroup}

\usepackage{graphicx}
\begin{document}


\begin{center}
{\Large
	{\sc 
	{Optimisation des parcours patients \\ 
	pour lutter contre l'errance de diagnostic \\ 
	des patients atteints de maladies rares \\}
	}
}
\bigskip

Frédéric Logé$^{1,2*}$, Rémi Besson$^{3*}$, Stéphanie Allassonnière$^{3}$ \\
\medskip
$^{1}$\it{Campus Innovation Paris, R\&D Air Liquide, 78350 Les Loges-en-Josas, France} \\
$^{2}$\it{Centre de Mathématiques Appliquées, Polytechnique, 91120 Palaiseau, France} \\
$^{3}$\it{Centre de Recherche des Cordeliers, Université de Paris, INSERM, Sorbonne Université, 75006 Paris, France} \\
\tt{frederic.logemunerel@gmail.com} ; \tt{remi.besson@polytechnique.edu}
\end{center}
\medskip
*Les auteurs ont contribué équitablement à la production de ce travail.
\bigskip


{\bf R\'esum\'e.} Un patient atteint d'une maladie rare en France doit en moyenne attendre deux ans avant d'être diagnostiqué. Cette errance médicale est fortement préjudiciable tant pour le système de santé que pour les patients dont la pathologie peut s'aggraver. Il existe pourtant un réseau performant de centres de référence maladies rares (CRMR), mais les patients ne sont orientés que trop tardivement vers ces structures. Nous considérons une modélisation probabiliste du parcours patient afin de créer un simulateur permettant d'entraîner un système d'alerte détectant les patients en errance et les orientant vers un CRMR tout en considérant les potentiels surcoûts associés à ces décisions. Les premiers résultats obtenus sur données simulées apparaissent prometteurs. Un important travail de mise en relation des données expertes disponibles avec les données de parcours patients reste à faire ainsi que des ajustements sur la modélisation proposée. \\
{\bf Mots-cl\'es.} Maladies rares, Parcours patient, Simulation, Graphes, Optimisation

\medskip

{\bf Abstract.} A patient suffering from a rare disease in France has to wait an average of two years before being diagnosed. This medical wandering is highly detrimental both for the health system and for patients whose pathology may worsen. There exists an efficient network of Centres of Reference for Rare Diseases (CRMR), but patients are often referred to these structures too late. We are considering a probabilistic modelling of the patient pathway in order to create a simulator that will allow us to create an alert system that detects wandering patients and refers them to a CRMR while considering the potential additional costs associated with these decisions. The first results obtained on simulated data appear promising. An important work of linking the available expert data with the data of patient journey remains to be done as well as adjustments on the proposed modeling. 
\bigskip \bigskip

\section{Introduction}
\subsection{Contexte}
Le récent rapport Erradiag\footnote{L'errance de diagnostic : Erradiag résultats de l'enquête sur l'errance de diagnostic. 2016. 844 patients interrogés.} de l'Alliance Maladies Rares a montré qu'en France un patient atteint d'une maladie rare\footnote{Maladie rare: maladie dont la prévalence est inférieure à 1/2000 (seuil Européen).} devra attendre en moyenne $2$ ans entre l'apparition des premiers symptômes et le diagnostic de sa pathologie. Pour plus d'un quart d'entre eux cette durée sera de plus de $5$ ans. Ce laps de temps, appelé errance de diagnostic, est un fléau pour le système de santé de par les surcoûts économiques (multiplication d'examens et traitements médicaux inutiles) et humain engendrés (aggravation de la pathologie du fait de l'inadéquation de la prise en charge).

Les patients atteints d'une maladie rare sont particulièrement affectés par l'errance de diagnostic du fait du caractère multifactoriel de ces pathologies : une maladie rare touche bien souvent plusieurs organes différents et le diagnostic nécessite donc une approche pluridisciplinaire. Par ailleurs, un praticien n'exerçant pas en centre expert n'observera probablement jamais la plupart des maladies rares au cours de sa carrière. Cependant, bien que les maladies rares soient par définition peu fréquentes, les malades sont eux nombreux puisque l'on estime à $3$ millions le nombre de personnes concernées en France et entre $263$ et $446$ millions dans le monde \cite{Nguengang}. Cela est dû au très grands nombre de maladies rares existantes : pas moins de $9000$ sont ainsi recensées dans OrphaNet\footnote{\url{https://www.orpha.net/consor/cgi-bin/index.php}. Accessed on [23/02/2020].}, le portail dédié aux maladies rares.     

La France est pionnière dans la lutte contre les maladies rares, lançant dès $2005$ le premier Plan National Maladie Rare avec notamment la structuration de centres de référence maladies rares (CRMR). Ces centres pluridisciplinaires ont pour but une meilleure prise en charge des patients atteints de maladies rares. Le rapport Erradiag rappelle que, malgré les progrès enregistrés suite à la création de ces centres, un nombre encore trop important de malades sont orientés très tardivement vers une structure hospitalière. Un quart des patients attendent ainsi plus de $4$ ans après les premiers symptômes avant que la recherche du diagnostic ne soit enfin initiée. Une meilleure orientation des patients dans le système de santé est donc nécessaire.

Dans le même temps les données médicales, sous forme de base experte comme Orphanet ou de données de parcours patient  (consultations/remboursements) avec l'apparition du dossier médical électronique, se sont structurées et sont désormais plus accessibles pour la recherche rendant possible l'objectif d'amélioration des parcours patient.

\subsection{Base expertes et données patients}

Il existe un certains nombre de base expertes pour les maladies rares, en particulier OrphaNet. Cette base référence plus de 9000 maladies rares, et leur associe les signes phénotypiques caractéristiques. C'est-à-dire que pour chaque maladie rare nous connaissons les symptômes généralement associés et nous connaissons la probabilité de présenter le symptôme sachant la maladie rare concernée. Cette base est inter-opérable avec HPO \cite{10.1093/nar/gkw1039} qui fournit un vocabulaire unifié sur les signes phénotypiques ainsi que l'ontologie associée. 

L'accès à des données de parcours patient est plus difficile puisqu'il s'agit de données personnelles. Le Plan Maladie Rare prévoit le déploiement de la Banque National De Maladies Rares, inter-opérable avec HPO et Orphanet, devant permettre de récolter des données sur les parcours des patients à l'intérieur des centres maladies rares. Cependant ces données ne fournissent pas, ou très partiellement, le parcours médical avant l'entrée dans un centre expert. 
    
Des données de type parcours patients consultation/remboursement (Electronic Health Record dans la littérature ou EHR) sont accessibles en ligne mais les données ne sont pas labelisées avec un identifiant Orpha pour les patients atteints de maladies rares. Certains travaux tels \cite{Colbaugh,Tremblay} contournent cette difficulté en annotant la base EHR à partir d'informations expertes (par exemple si tel médicament a été pris il s'agit probablement de telle maladie rare) mais une telle approche ne peut qu'être limitée à certaines maladies rares.

Le lien entre les bases de données expertes du type OrphaNet et les bases de données de parcours patient n'est pas aisé à opérer. En effet, alors qu'OrphaNet associe des symptômes (identifiants HPO) à des maladies rares (identifiants Orpha), les EHR témoignent du parcours de santé (remboursements médicament, consultation généraliste/spécialiste). Un premier défi est d'associer à un évènement de santé d'EHR un symptôme HPO. \cite{zhang2019unsupervised} propose ainsi une approche de type NLP pour annoter le texte libre d'EHR avec des identifiants HPO. Nous faisons ici l'hypothèse qu'un tel lien est possible et qu'il est ainsi possible de combiner des données expertes avec des données cliniques ce qui est essentiel pour le cas des maladies rares où les données cliniques seules ne suffisent pas mais permettent d'introduire une notion de temporalité absente d'Orphanet. 

\subsection{Objectif}

Notre but ici est de montrer comment nous pouvons améliorer les parcours patients en réduisant l'errance de diagnostic pour les individus atteints de maladies rares. En particulier, nous nous focalisons sur un système d'alerte, lequel suggère à un patient de se rendre en centre expert avec l'information du diagnostic suspecté et des anomalies à rechercher.

Dans la section \ref{sec:modelisation} nous présentons notre modélisation mathématique du problème de prise de décision : envoyer ou non un patient en CRMR. Dans la section \ref{sec:resultats} nous présentons l'application de notre approche sur des parcours patients simulés, dont le code R est disponible en libre accès. Nous enchaînons ensuite avec la discussion des résultats et une conclusion générale.

\section{Modélisation}\label{sec:modelisation}

\subsection{Parcours patient}

Nous considérons qu'un individu ne peut contracter qu'un seul syndrome à la fois. Ce patient ne suspecte la présence de ce syndrome que via l'apparition de symptômes. Nous notons alors $S_i$, $i \in \mathbb{N}$, l'ensemble des symptômes observés le $i$-ème jour suivant l'apparition de la maladie et $H_t = \{ S_i ; 1 \leq i < t \}$ l'historique de ces observations jusqu'au jour $t-1$. Un exemple de parcours patient est présenté dans la figure \ref{fig:main_results}, graphe de gauche.

\subsection{Symptômes}

Nous supposons que les symptômes se distribuent en trois groupes :
\begin{itemize}
\item Les symptômes latents, qui ne sont pas observables sans examen médical approprié. Exemple : malformations cardiaques ou neurologiques.
\item Les symptômes visibles de façon permanente. Exemple: malformation externe.
\item Les symptômes récurrents et passagers. Il s'agit de symptômes pouvant apparaître de manière récurrente puis disparaître et apparaître de nouveau. Exemple: migraines, otites.
\end{itemize}

Pour cette étude nous avons simulé\footnote{Les détails de la simulation sont fournis à l'adresse \url{github.com/FredericLoge/patientPathway}} un graphe faisant le lien entre syndromes et symptômes, ainsi que la probabilité d'occurrence d'un symptôme dans la durée, comme représenté figure \ref{fig:graphe_syndrome_symptomes}, graphe de gauche. Nous avons supposé que les temps d'occurrence suivaient une loi gaussienne, tronquée à gauche par 0. Pour les symptômes de type chronique, nous avons supposé que les délais entre présence/absence des symptômes suivaient une loi exponentielle.

La base OrphaNet nous fournit la prévalence d'une maladie rare ainsi que la probabilité des signes phénotypiques associés. En combinant connaissances expert et données de parcours patient nous pourrons calibrer des modèles paramétriques du délai d'apparition des symptômes sachant les syndromes.



\subsection{Prise de décision : envoi en CRMR}\label{sec:definition_cout}

Basé sur l'historique de l'individu, et suite à l'apparition d'au moins un symptôme, nous allons vérifier s'il est pertinent pour le patient de se diriger vers un centre spécialisé en maladie rare.

Nous estimons un prédicteur $\hat{f}$ de la probabilité qu'un individu soit atteint d'une maladie rare. Ce prédicteur prend en entrée l'historique $H_t$ précédemment décrit. A partir de l'apparition du premier symptôme, nous vérifions chaque jour si, oui ou non, cette probabilité dépasse un seuil $\tau, \tau \in [0,1],$ préalablement fixé. Si le seuil est dépassé, le patient est envoyé en CRMR et son syndrome, rare ou non, sera découvert. La sollicitation des médecins spécialisés et les tests réalisés coûteront un certain prix. Si le seuil n'est pas dépassé, le patient ne sera pas envoyé en CRMR et nous considérons pour la modélisation qu'il est envoyé à la fin de la période de temps d'analyse. Nous notons cette procédure $\pi_{\tau}$ qui à un historique $H_t$ associe la décision binaire $\mathbb{1}\{ \hat{f}(H_t) > \tau \}$. 

Soit les évènements $E = \{$Le patient a une maladie rare$\}$ et $A = \{$Nous l'avons dirigé vers un centre de maladie rare$\}$. Nous pouvons alors définir un coût pour un parcours patient selon ces deux évènements :
\begin{equation}\label{eq:definition_cout}
\cout = 
\begin{cases}
     \coutErranceParJour \cdot \tempsErrance & \\
     \qquad + \coutMedecinSpeMR & \text{si } {E} \cap {A}  \\
     \coutErranceParJour \cdot \tempsMoyenErrance & \\ 
     \qquad + \coutMedecinNonSpeMR \cdot \nbMoyenMedecinConsultes & \text{si } {E} \cap \bar{A}  \\
     \coutErranceParJour \cdot \tempsErrance & \\
     \qquad + \coutMedecinSpeMR & \text{si } \bar{E} \cap {A}  \\ 
     \coutErranceParJour \cdot \tempsErrance & \\
     \qquad + \coutMedecinNonSpeMR & \text{si } \bar{E} \cap \bar{A} 
\end{cases}
\end{equation}
où $\tempsErrance$ corresponds au temps entre l'observation du premier symptôme et la prise de décision (si prise de décision, sinon dernier jour de la simulation). Les coûts détaillés peuvent être recueillis auprès d'experts du domaine médical.

Notre objectif est de déterminer, pour un prédicteur de maladie rares $\hat{f}$ le seuil $\tau$ approprié pour optimiser le coût introduit dans l'équation \ref{eq:definition_cout}. Formellement, nous cherchons à résoudre
\begin{equation}\label{eq:fonction_objectif}
\pi^* = \arg\,\min_{\tau \in [0,1]} \mathbb{E}_{\syndrome \sim \nu}\left[ \cout \mid \syndrome \right]
\end{equation}
où $\nu$ est la probabilité de distribution sur les syndromes.

\section{Résultats}\label{sec:resultats}

Tous les détails sur les résultats et simulations présentées ici peuvent être obtenus à l'adresse \url{github.com/FredericLoge/patientPathway}. Des graines ont été soigneusement établies dans le code pour s'assurer de sa reproductibilité.

\begin{figure}
    \centering
    \vcenteredhbox{\includegraphics[width = 7cm]{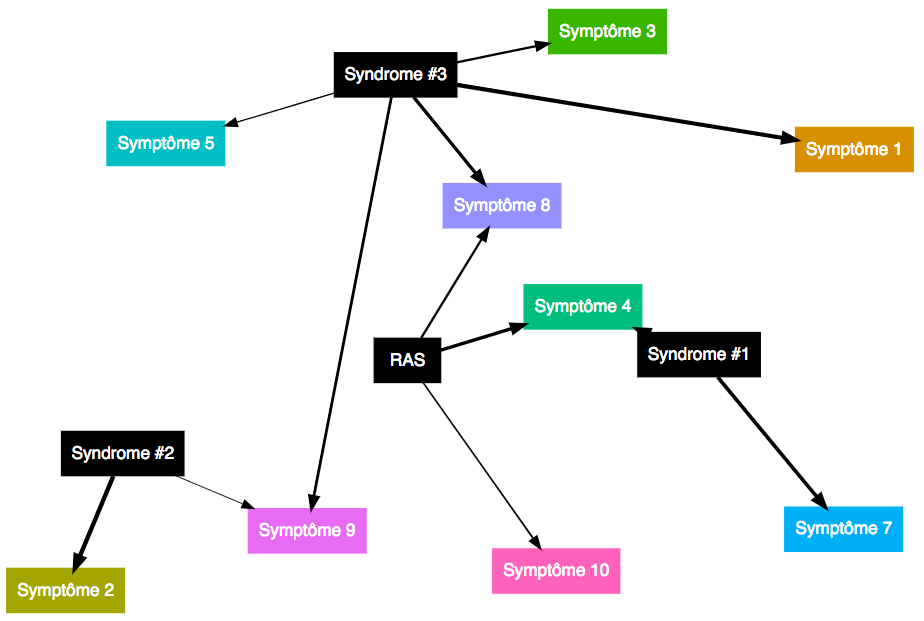}}
    \hspace{0.5cm}
    \vcenteredhbox{\includegraphics[width = 7cm]{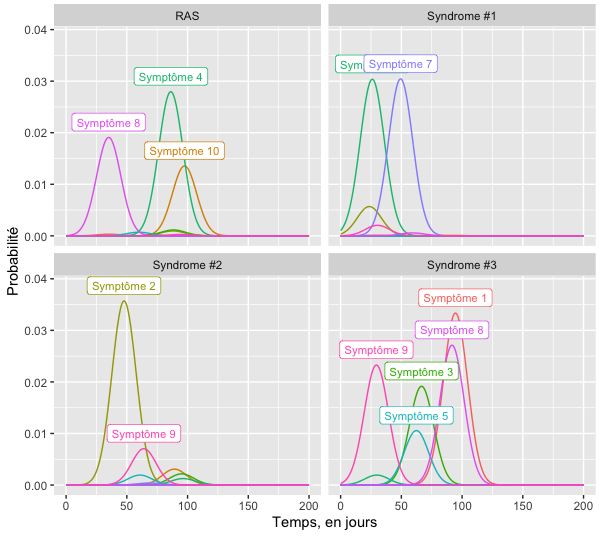}}
    \caption{\textit{(Gauche)} Relations entre symptômes syndromes considérés, l'épaisseur du trait indiquant la probabilité qu'un symptôme se déclare lorsque le patient est atteint d'un syndrome. Note : les liens dont la probabilité étant inférieure à 15\% ont été exclus pour une meilleure lisibilité. \textit{(Droite)} Probabilité d'occurrence, dans le temps, des différents symptômes conditionnellement au syndrome considéré, le jour 0 étant l'instant où le syndrome se met au place.}
    \label{fig:graphe_syndrome_symptomes}
\end{figure}

Pour faciliter l'étude, nous avons simulé un graphe symptômes-syndromes ainsi que les probabilités d'occurrence dans le temps des différents symptômes comme représenté dans la figure \ref{fig:graphe_syndrome_symptomes}. Quatre syndromes ont été considérés dont un RAS (Rien à Signaler) et un appartenant à la catégorie des maladies rares, le syndrome \#1. Un total de dix symptômes a été considéré.

\begin{figure}
    \centering
    \vcenteredhbox{\includegraphics[height = 6cm]{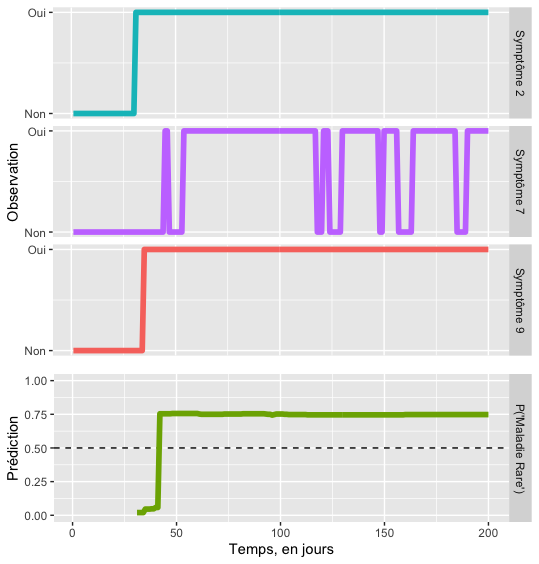}}
    \hspace{0.5cm}
    \vcenteredhbox{\includegraphics[height = 6cm]{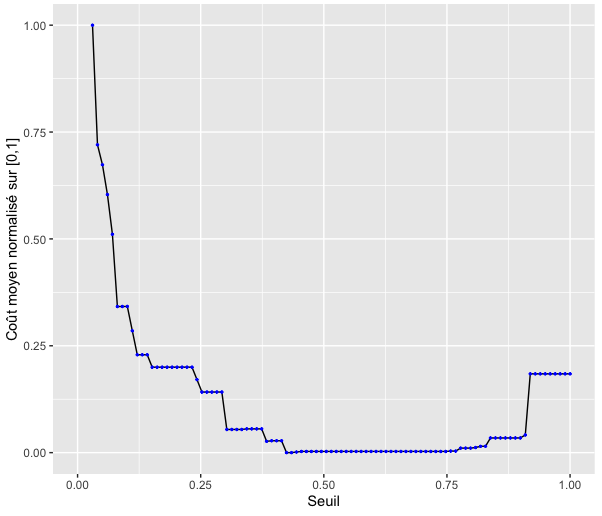}}
    \caption{\textit{(Gauche)} Données simulées sur 200 jours, patient atteint du syndrome \#1 (maladie rare pour rappel) qui donne souvent lieu aux symptômes 4 et 7, ici les symptômes 2 puis 9 et finalement 7 vont être observés (trois premières courbes), le symptôme 4 est présent mais non observé car il s'agit d'un symptôme latent. La prédiction du risque de maladie rare, opérée dès l'observation du premier symptôme, le symptôme 2, franchit le seuil de 0.5 dès l'observation du symptôme 7.
    \textit{(Droite)}. En simulant les données comme représentées à gauche pour les différents syndromes et un grand nombre de fois, nous avons pu établir le coût moyen, comme défini dans la section \ref{sec:definition_cout} en fonction du seuil utilisé pour prendre la décision d'envoyer la personne en centre de maladie rare. Ce coût est ici représenté sur l'échelle [0;1]. A l'extrême gauche, nous avons le coût associé au fait d'envoyer tous les patients en centre spécialisé directement. A l'extrême droite, nous avons le coût associé au fait de n'envoyer aucun patient en centre spécialisé.}
    \label{fig:main_results}
\end{figure}

Nous avons généré pour chaque syndrome 100 parcours patient sur une durée de quatre ans, à pas de temps journalier. Ces données ont été utilisées pour calibrer une forêt aléatoire (voir \cite{hastie2009elements} chapitre 15)
prédisant la probabilité que le patient suivi ait une maladie rare en se basant sur les symptômes observés et la temporalité de leurs apparitions relativement au premier symptôme. Une fois que le prédicteur est entraîné, nous pouvons estimer la fonction objectif exprimée dans l'équation \ref{eq:fonction_objectif} sur une grille de $\tau$ assez fine, comme représenté sur la figure \ref{fig:main_results}. Sur le graphe de gauche, la courbe verte indique la prédiction en fonction du temps, cette prédiction est utilisée pour notre système d'alarme. L'observation temporelle des symptômes 2, 7 et 9 sont également représentés. Sur le graphe de droite on observe le coût moyen en fonction du seuil choisi. Comme attendu, il y a un équilibre à trouver entre envoyer tout le monde en CRMR ($\tau = 0$, extrême-gauche du graphe), et n'envoyer personne ($\tau = 1$, extrême-droite du graphe).

\newpage 
\section{Conclusion et discussion}

En France, un patient souffrant d’une maladie rare doit en moyenne attendre deux ans avant d’être diagnostiqué. A cette errance médicale sont associés des coûts économiques et humains élevés alors même que des centres experts maladies rares ont été déployés sur le territoire et donnent satisfaction lorsqu'ils sont sollicités. 

Nous avons donc proposé ici une procédure pour la création d'un système d'alerte permettant de détecter dans des bases de données de parcours patients les patients atteints de maladie rare et devant être orientés vers des centres experts. Nous prenons en compte dans ces décisions les différents coûts engendrés par les diverses actions possibles. Ce système d'alerte est entraîné en générant des données issues d'un simulateur de parcours patient que nous construisons à partir de données expertes et dont les paramètres devront être calibrés avec des données cliniques. Les premiers résultats, obtenus sur données simulées, apparaissent prometteurs. Toutefois un important travail doit être encore être fourni sur les points suivants :

\paragraph{Calibrer le modèle Symptômes-Syndrome} Ce modèle doit être, à termes, calibré à partir d'un mélange de données expertes et de parcours patients. Ces données peuvent être combinées au prix d'un travail significatif. Il s'agit de constituer un module associant à chaque évènement de santé d'un EHR un identifiant HPO. Une annotation experte doit également permettre de mieux modéliser l'apparition des symptômes au cours du temps : information sur l'âge typique d'apparition d'un symptôme, les durées typiques, la nature du symptôme.


\paragraph{Améliorer et valider la modélisation} Dans notre modèle de parcours patient, le médecin n'était présent qu'une fois la décision de se rendre en CRMR prise ou la durée de simulation terminée. Dans la réalité, l'occurence d'un symptôme incitera probablement la personne à consulter une personne du corps médical. Ces agents doivent être impérativement intégrés au parcours patient. Des approches de validation par indicateurs comme celle développé dans \cite{prodel:tel-01665163} seront utilisées.

\paragraph{Le problème de contrôle} Nous avons considéré ici une stratégie particulière pour la réorientation vers les centres experts qui exécute un choix binaire si l'on dépasse un certain seuil de probabilité que le patient présente une maladie rare. Une future direction est de formuler le problème par un processus décisionnel de Markov où la stratégie pourra être apprise par des algorithmes d'apprentissage par renforcement se basant sur des données du simulateur et n'ayant plus nécessairement une forme aussi spécifique.

\paragraph{Remerciements :} Ce travail a bénéficié d'une aide de l’État gérée par l'Agence Nationale de la Recherche au titre du programme d’Investissements d’avenir portant la référence ANR-19-P3IA-0001.

\bibliographystyle{apalike}
\bibliography{refs}

\end{document}